\documentstyle[12pt]{article}

\begin{document}

\baselineskip=24pt

\def \HI{H~{\small I}}
\def \HII{H~{\small II}}
\def \HA{H$\alpha$}
\def \deg{$^{\circ}$}
\def \BT0{$B_{\rm T}^{0}$}
\def \VT0{$V_{\rm T}^{0}$}
\def \BVT0{$(B-V)_{\rm T}^{0}$}
\def \UBT0{$(U-B)_{\rm T}^{0}$}
\def \LB{$L_{\rm B}$}
\def \fB{$f_{\rm B}$}
\def \fBfV{$f_{\rm B}/f_{\rm V}$}
\def \LFIR{$L_{\rm FIR}$}
\def \fFIR{$f_{\rm FIR}$}
\def \fsix{$f_{60}$}
\def \ften{$f_{100}$}
\def \Lsolar{$L_{\odot}$}
\def \Msolar{$M_{\odot}$}
\def \fFIRfB{$f_{\rm FIR}/f_{\rm B}$}
\def \LFIRLB{$L_{\rm FIR}/L_{\rm B}$}
\def \fsixfB{$f_{60}/f_{\rm B}$}
\def \fHIfB{$f_{\rm H~{\small I}}/f_{\rm B}$}
\def \ftenfsix{$f_{100}/f_{60}$}
\def \fsixften{$f_{60}/f_{100}$}
\def \12COJ{$^{12}$CO ($J$ = 1 - 0)}

\title{
A Variation of the Present Star Formation Activity
of Spiral Galaxies
}

\author{ 
\bigskip
Akihiko T{\small OMITA},
Yoshio T{\small OMITA}, 
and Mamoru S{\small AIT\={O}}
\\
{\it
Department of Astronomy,
Faculty of Science, 
Kyoto University,
}
\\
{\it
Sakyo-ku, Kyoto 606-01
}
\\
{\it
E-mail (AT)
atomita@kusastro.kyoto-u.ac.jp
}
}

\date{
(Received 1995 November 7; accepted 1996 January 5)
\bigskip
\\
To be appeared in PASJ, Vol. 48, No. 2 (April 1996 issue)
}

\maketitle

\newpage

\section*{Abstract}

\indent

The star formation rate in spiral galaxies is considered
to be decreasing continuously with time
in a time scale of $10^{9}$ yr.
The present star formation activity,
on the other hand,
occurs in molecular clouds
with a time scale of $10^{7}$ yr, 
and shows various degrees among galaxies.
We make a new data set of 1681 nearby spiral galaxies
from available databases
and study the statistics of the present star formation activity.
We analyze far-infrared and optical $B$-band surface brightnesses
of the \HII \ regions and the non-\HII \ regions in M~31
and show that a far-infrared-to-optical $B$-band flux ratio, \fFIRfB,
is a useful indicator of the present star formation activity
of spiral galaxies.
For the sample galaxies,
we make the distribution diagram of log~(\fFIRfB) versus log~\LB \
for each morphological type.
The distribution of \fFIRfB \ has a dispersion
of one to two orders of magnitude even within the same
morphological type of galaxies,
implying that the star formation activities of spiral galaxies
changes discontinuously in a short time scale.
Analyzing the log~(\fFIRfB) versus log~\LB \ correlation,
we suggest that the most active star formation in galaxies
does not continue longer than $10^{8}$~yr.
We also construct a universal distribution histogram
of log~(\fFIRfB) for
each morphological type.
The earlier-type spirals tend to show larger variation of
the present star formation activity.
We discuss the correlation between the present
star formation activity
and the structures and environments of galaxies.
We suggest that the short-term variation occurs primarily
due to internal processes
which may change with the morphological type of galaxies.

{\bf Key words:}~
Galaxies: star formation ---
Galaxies: evolution ---
Galaxies: star formation indicators ---
Galaxies: morphological types

\newpage

\section{Introduction}

\indent

The star formation rate (SFR) in spiral galaxies
with a time scale of $10^{9}$ yr seems to be
continuously decreasing with time.
The late-type spirals have reduced the SFR
with a time scale comparable to the Hubble time
(e.g., Searle et al. 1973; Kennicutt 1983; Gallagher et al. 1984),
while the early-type spirals have
a low preset SFR compared with the past SFR
and the decaying time scale is much shorter than 
the Hubble time scale (e.g., Kennicutt 1983, Sandage 1986).
Photometric evolution models
under such a view of the star formation histories
reproduced the present colors of galaxies along the Hubble sequence
(e.g., Bruzual 1983; Arimoto, Yoshii 1986;
Guiderdoni, Rocca-Volmerange 1987).
Stars, especially massive stars,
mostly form in giant molecular clouds (GMC).
The time scale of the star formation activity in a GMC
is of order of $10^{7}$ yr (e.g., Shu et al. 1987),
and the dynamical time scale of a galaxy is $\sim$$10^{8}$~yr;
these are much shorter than the time scale of Gyr.
In the galaxy scale,
some galaxies show transient starburst phenomenon
and there are post-starburst galaxies
(e.g., Walker et al. 1988; Beckman et al. 1991).
These suggest that the active star formation occurs discontinuously
in a short time scale.
This has been also suggested by statistical studies
of far-infrared (FIR)-to-optical
$B$-band flux ratios of galaxies
(e.g., de Jong et al. 1984; Soifer et al. 1987; Bothun et al. 1989).

It is also a controversial question what kind of global parameters
of galaxies are closely related to
the present star formation activity.
Roberts et al. (1979) and Kenney et al. (1992)
argued that the bar structure triggers a nuclear starburst.
However, Moss and Whittle (1993) showed from analysis of 
the \HA \ equivalent width that
the existence of bar is not correlated with
the present star formation activity.
Dressler et al. (1985) claimed that cluster galaxies
have smaller star formation activity than field galaxies,
while Gavazzi and Jaffe (1985) and Moss and Whittle (1993)
concluded that cluster galaxies have more intense star formation
activity.
The galaxy-galaxy interaction is considered to be an
effective triggering mechanism for starburst (e.g., Keel 1993),
while Campos-Aguilar and Moles (1991) suggested that occurrence of
the starburst is independent of the galaxy-galaxy interaction.

The best way to study the star formation features in a short time scale
and its connection with the global characteristics of galaxies is
to make a reliable indicator of
the present star formation activity and
to investigate the difference of the present star formation activity
among various galaxies.
In this paper,
we construct a new large high-quality data set 
of spiral galaxies using available databases.
We describe the sample in section~2.
In section~3
we make a far-infrared (FIR)-to-optical $B$-band flux ratio,
\fFIRfB, for each galaxy and
show a systematic distribution of the flux ratios
of the sample galaxies.
Using the measured FIR and $B$ band surface brightnesses
at various areas of M~31,
we confirm that the \fFIRfB \ is a useful measure of
the present star formation activity in the galaxy scale.
We present for the first time
a distribution of log~(\fFIRfB) of spiral galaxies
for each morphological type.
The distributions of \fFIRfB \ show that
the present star formation activity has
a great variety even within the same morphological type of galaxies
and suggest that the star formation activity in a short time scale
changes discontinuously.
In section~4 we discuss
what kind of global features of galaxies are
correlated to the present star formation activity.
A summary is given in section~5.

\section{The Sample}

\indent

We are interested in the star formation histories of spiral galaxies
and select a sample of spirals using
{\it Third Reference Catalogue of Bright Galaxies}
(de Vaucouleurs et al. 1991; hereafter RC3).
RC3 is the largest optical galaxy catalog
describing various parameters of galaxies.
We made use of {\it NAO/ADAC CD-ROM Catalog Volume~1}
to search the RC3 data.

The $B$ magnitudes of galaxies provide
a measure of the galaxies
and nearly represent the total mass (e.g., Roberts, Haynes 1994).
RC3 gives an extinction-corrected $B$ magnitude
of each galaxy which is denotes as \BT0
and corresponds to $B$ magnitude as is observed at face-on.
In the following analyses,
we also use the data of the color \BVT0,
the morphological index $T$,
and the radial velocity in RC3;
the optical color of the galaxy is considered to be a result
of the star formation history in the galaxy
(e.g., Larson, Tinsley 1978; Arimoto, Yoshii 1986;
Guiderdoni, Rocca-Volmerange 1987),
the morphological type is one of the fundamental parameters
of galaxies (e.g., Roberts, Haynes 1994),
and the radial velocity is used to estimate the distance and
the luminosity of the galaxy.
Out of 13537 galaxies with the morphological index
of $0 \leq T \leq 11$ (spirals and irregulars) cataloged in RC3,
1782 have all values of
\BT0, \BVT0, and radial velocity.
We select galaxies regardless of the bar structure.
We also use the far-infrared (FIR) emission fluxes of galaxies
measured by $IRAS$ as an essential parameter of this study.
Out of the 1782 galaxies,
101 lie in the non-surveyed regions of $IRAS$ mission
or belong to close binary systems
for which we do not identify the association of the $IRAS$ source.
We thus finally adopt 1681 galaxies.

Table~1 shows the morphology and magnitude distribution
of the sample galaxies
and table~2 shows the morphology and radial velocity distribution;
about 80\% of the sample galaxies are brighter than 14~mag and
have redshifts less than 5000~km~s$^{-1}$.
The sample fairly covers all of the morphological types.
Figure~1 shows a distribution histogram
of \BVT0 for each morphological type.
There is a correlation between the \BVT0 color and the morphology;
the later the morphological type is,
the bluer the \BVT0 color is.
The variation from S0a to Sc is larger than that among later types.
The difference of the median of the \BVT0 color
between the adjacent morphological types is comparable to
the dispersion of the \BVT0 color in each morphological type,
$\Delta$~\BVT0~$\sim$ 0.3.
These color characteristics along the spiral Hubble sequence have
been known (e.g., Sauvage, Thuan 1994; Roberts, Haynes 1994).
Our sample also has the same trend.

\section{Star Formation Activity in a Short Time Scale}

\subsection{An Indicator of Star Formation Activity
in Spiral Galaxies}

\indent

It is well known that the \HA \ equivalent width
is a direct indicator of the present star formation rate
(e.g., Kennicutt 1983).
However,
the \HA \ emission suffers from the extinction by
interstellar medium (ISM)
and it is difficult to obtain the total \HA \ flux of the galaxy
(Kennicutt 1992).
The FIR emission of spiral galaxies
originates from both warm dust heated by massive stars
and cool dust heated by general stellar field
(Presson, Helou 1987; Devereux, Young 1990; Sauvage, Thuan 1992;
Devereux, Scowen 1994; Devereux et al. 1994; Devereux et al. 1995).
Gavazzi et al. (1991) showed that the FIR flux normalized to optical
flux has a correlation with the \HA \ equivalent width
in two orders of magnitude.
Xu (1990) suggested that radiation from dust heated by
UV continuum dominates the total FIR emission of spiral galaxies.
These indicate that the FIR flux is also an indicator of
the present star formation rate in spiral galaxies.
The FIR-to-$B$ flux ratio has been used as an indicator for
the present star formation activity in galaxies
(de Jong et al. 1984; Fabbiano et al. 1988;
Soifer et al. 1987; Bothun et al. 1989; Sauvage, Thuan 1994).
Belfort et al. (1987) also showed from their photometric synthesis model
that the FIR-to-$B$ flux ratio responds to a recent starburst
within 2 $\times$ $10^{7}$ yr.

Following Soifer et al. (1987),
we define the $B$-band flux \fB \ 
as 4400~\AA \ times flux density [W~m$^{-2}$~\AA$^{-1}$] at 4400~\AA.
Using an equation between the $B$-band flux and the magnitude system
given in Allen (1973),
we obtain log~\fB \ [W~m$^{-2}$] = $-$0.4~\BT0 $-$ 7.53.

The $IRAS$ survey has provided us FIR fluxes for
a large, homogeneous, nearly all-sky, flux-limited sample,
which are free from the heavy extinction by the ISM.
The total FIR fluxes for galaxies are available
using following five $IRAS$ catalogs.
The FIR fluxes of the most sample galaxies are taken from
{\it IRAS Faint Source Catalog} (FSC) and
{\it IRAS Point Source Catalog} (PSC).
Since the measured aperture of $IRAS$ point sources is about 2~arcmin,
we underestimate the FIR fluxes for
the galaxies with angular sizes larger than 2~arcmin.
We refer to {\it IRAS Small Scale Structure Catalog} (SSSC)
for the galaxies with sizes of about 2 to 8~arcmin.
For the galaxies with sizes larger than 8~arcmin,
we adopt the data of Rice et al. (1988).
For the galaxies having 60~$\mu$m flux densities greater than 5.24~Jy,
we take the data of the FIR-bright galaxies by Soifer et al. (1989).
Cross identification with optical objects
was performed mainly based on the description
in the above-mentioned catalogs and 
{\it Cataloged Galaxies and Quasars in the IRAS Survey}
(hereafter CGQIRAS)
and it was also checked using NASA/IPAC Extragalactic Database (NED).
We made use of {\it ADC CD-ROM Selected Astronomical Catalogs Volume 1}
in searching the data of PSC, FSC, SSSC, and CGQIRAS.
We calculate
\footnote{
The readers would notice the dimension discrepancies
in these equations.
These are only arithmetical formulae.
}
the FIR flux from the flux densities at 60 and 100~$\mu$m,
\fsix \ and \ften,
following CGQIRAS;
i.e., log~\fFIR \ (42.5~-~122.5~$\mu$m) [W~m$^{-2}$]
= log~(2.58~\fsix~[Jy]~+~\ften~[Jy]) $-$~13.90.
1150 galaxies have reliable measurements in both \fsix \ and \ften.
In figure~2,
we obtain an empirical formula$^{1}$,
log~(\ftenfsix) =
$-$0.23~log~(\fsix~[Jy]/\fB~[W~m$^{-2}$]) +~3.41
with a correlation coefficient of 0.72,
from 1150 galaxies with reliable \fsix \ and \ften.
Objects with relatively large deviations are mostly late-type spirals
or irregulars;
a correlation coefficient using only objects with
0 $\geq$ $T$ $\geq$ 5 is 0.80.
Using the regression line,
we estimate \ften \ for 156 objects with upper-limited \ften \
and reliable \fsix,
and calculate \fFIR \ of these objects.
The FIR fluxes estimated by this method have an error of 4\%
in log~(\ftenfsix),
which is negligible in the following analyses.
An upper limit of log \fFIR \ = $-$13.82 [W~m$^{-2}$]
is given for 375 sample galaxies without $IRAS$ detection
adopting the detection limits of $IRAS$ survey of
\fsix \ = 0.2~Jy and \ften \ = 0.7~Jy (FSC).

\subsection{\fFIRfB \ vs \ftenfsix \ Characteristics
at Various Regions of M31}

\indent

Figure~3 shows a log~(\ftenfsix) versus log~(\fFIRfB) relation
for the 1150 sample galaxies
which have reliable \fsix \ and \ften \ values.
The color temperatures corresponding to log~(\ftenfsix)
shown in the ordinate is about 41.7~/~[1~+~log~(\ftenfsix)]~K
for a modified black-body radiation in the form of
$\nu^{1.5}B_{\nu}$,
which gives about 42~K, 30~K, and 23~K for log~(\ftenfsix)~=
0, 0.4, and 0.8, respectively.
The values of \fFIRfB \ shown in the abscissa
are distributed over two orders of magnitude and
the galaxies with higher values of \fFIRfB \ tend to have
smaller \ftenfsix \ values,
i.e., higher temperatures;
the correlation coefficient is 0.63 for the total 1150 sample
and 0.72 for 916 objects with 0 $\geq$ $T$ $\geq$ 5,
shown by larger circles in figure~3.
The log~(\ftenfsix) versus log~(\fsixfB) relation has
a somewhat tighter correlation
(see section~3.1 and figure~2),
because \fsix \ is more insensitive for the ``cirrus'' component
than \ften \ (e.g., Presson, Helou 1987).
In figure~3,
six famous spiral galaxies without strong AGN are denoted 
by filled squares with numbers,
1: M~31, 2: M~81, 3: M~33, 4: M~101, 5: M~51, and 6: M~83.

de Jong et al. (1984) presented that
the values of the \fFIRfB \ of galaxies are distributed
in two orders of magnitude
by analyzing 165 optically bright galaxies.
Soifer et al. (1987) showed that the \fFIRfB \ of
313 FIR-bright galaxies are independent of
the $B$ luminosity over two orders of magnitude,
and suggested that the star formation activity of galaxies
considerably varies in a short time scale.
Bothun et al. (1989) confirmed that
the values of \fFIRfB \ distribute over
two orders of magnitude among about 3500 UGC galaxies.
However, Bothun et al. (1989) also claimed that
it is dangerous to take the \fFIRfB \ value
as a direct indicator of the present star formation activity because
``cirrus'' component may contribute much in the FIR emission.
We examine the usefulness of the \fFIRfB \ value
as an indicator of the present star formation activity
using the data of M~31.

Rice (1993) made {\it IRAS Nearby Galaxy High Resolution Image Atlas}
(hereafter {\it IRAS HiRes}) of 30 galaxies
for $IRAS$ four bands with a 1~arcmin resolution,
using an image reconstruction technique;
among them there are 2\deg $\times$ 2\deg \ (481 $\times$ 481 pixels)
FITS images of M~31 ($T$ = 3.0).
Because of the proximity of M~31,
their $IRAS$ images can separate each of giant \HII \ regions.
The maps are distributed as
Infrared Processing and Analysis Center (IPAC)
released products.
We made use of the on-line service via
Automated Retrieval Mail System (ARMS) at
NASA's Data Archive Distribution Service (NDADS).
One of us (Y.~T.) and H. Suzuki obtained a $B$-band image of M~31
at Ouda Station, Kyoto University using
the Wide Field CCD Camera of fl = 300~mm and F/4.5;
its field of view is 2.\deg 5 $\times$ 1.\deg 7
(576 $\times$ 384 pixels).
The observation was made at October 27, 1992 and the total exposure time
was 20 min.
The sky subtraction was made with SPIRAL (Hamabe, Ichikawa 1992)
and foreground stars were removed by using an IRAF
\footnote{
IRAF is the software developed in National Optical Astronomy
Observatories, USA.}
task, `imedit'.
The $B$-band image of M~31 is shown in figure~4~(a).
Figure~4~(b) shows \fFIRfB \ image of M~31,
made by using the $B$-band image and the FIR image
constructed from \fsix \ and \ften \ images of {\it IRAS HiRes}.
From the \fFIRfB \ and \ftenfsix \ ratio maps,
we obtain the ratios
at some areas of
\HII \ regions in disk (2 to 8~kpc),
non-\HII \ regions in disk (2 to 8~kpc),
bulge region (1 to 2~kpc), and
nuclear region (within 0.3~kpc),
where the distances in parentheses are measured
from the nucleus of M~31
and the aperture radius is 0.$'$5.
The sample locations are shown on
the $B$-band and \fFIRfB \ images of figure~4.

Figure~5 shows the log~(\fFIRfB) versus log~(\ftenfsix) relation
for the various locations in M~31,
where the extinction in $B$ band was assumed to be
$\Delta$ log \fB \ = 0.4.
Both the non-\HII \ regions in the disk and the bulge region
are located at similar positions on this diagram and
we simply call both of them as the non-\HII \ regions,
which correspond to the cirrus region.
The plots in figure~5 are concentrated into three regions,
which are
the \HII \ regions (A),
the non-\HII \ regions (B), and
the central regions (C).
A big star mark in figure~5 indicates the whole of M~31.
The three regions and the whole M~31 are also
superimposed on figure~3
which shows the log~(\fFIRfB) versus log~(\ftenfsix) relation
of the sample galaxies.
In M~31,
when the aperture size is increased outward the \HII \ regions,
the values of \ftenfsix \ and \fFIRfB \ move on the diagram of figure~5
from the region~A to the region~B.
A solid line in figure~3 indicates the path of the plots
generated by increasing the aperture size.

In figure~5,
the nuclear region occupies different positions
from the \HII \ region and the non-\HII \ region positions.
Only 4\% of the total FIR emission comes from the nuclear region
in M~31 (Devereux et al. 1994),
and the nuclear region has little contribution
to the colors of the whole of M~31.
There are a small number of galaxies among our sample
whose FIR emissions are heavily contaminated
by AGN.
We referred to NED to search for the objects with AGN.
We found that only 3.6\% (34 objects) of our total sample are objects
with AGN including possible and suspect ones.
Most of AGNs are LINERs or type-2 Seyferts, i.e., weak AGNs.
Even for M~81,
a galaxy with LINER,
the nuclear region emits only 17\% of the total FIR emission
(Devereux et al. 1995).
Though we did not exclude the objects with AGNs from our sample,
the AGN-contamination affects little on the following analyses.

We find in figure~3 that
galaxies with the smallest \fFIRfB \ 
are located around the non-\HII \ region (B) area,
and that the galaxy distribution on the diagram is along
the path shown by the solid line.
In figure~3,
the active star-forming galaxies,
M~101 (No.~4), M~51 (No.~5), and M~83 (No.~6) lie in the region~A
and the inactive star-forming galaxy M~31 (No.~1) lies in the region~B.
The order from No.~1 (M~31) to No.~6 (M~83) represents the
magnitude of the star formation activities usually accepted for
those galaxies (e.g., Kennicutt, Kent 1983; Kennicutt et al.  1989;
Kennicutt et al. 1994).
Beichman (1987) made an $f_{25}/f_{60}$ versus \fsixften \
diagram of galaxies and showed that
galaxies on this color-color diagram are located
between the cold cirrus region and the hot \HII \ region,
consistent with our analysis in M~31.
The spatial resolution of 1~arcmin in {\it IRAS HiRes} corresponds to 
about 200~pc on M~31,
which is a little larger than typical \HII \ region size of
dozens to hundred~pc.
Therefore,
the values of \fFIRfB \ derived for the \HII \ regions are
somewhat smaller than that the true values of the \HII \ regions
and the true \HII \ regions may move toward lower-right
on the diagram of figure~5.
The \HII \ regions in M~31 do not contain early O-type stars
(Devereux et al. 1994).
Wood and Churchwell (1989) found that ultra-compact \HII \ regions
in the Galaxy have log~(\ftenfsix)~= 0 to 0.5,
which are similar to those of the region~A of M~31.
Thus,
the region~A of M~31 represents the position of universal
\HII \ region on the diagram.
Moreover,
most of the $IRAS$ sources associated with low-mass
young stellar objects have values of log~(\ftenfsix)
similar to those of the region~A (e.g., Beichman et al. 1986).
The region~A seems to represent the position of
the star-forming region in general.

We thus conclude that \fFIRfB \ is
a useful indicator of the present star formation activity
in the whole galaxy,
although the influence of the cirrus component is contained
in the value of \fFIR.
We show in appendix~1 that \fFIRfB \ is not affected
by the internal absorption of galaxy
by checking a correlation between the distribution of log~(\fFIRfB)
and apparent inclination of galaxies.
It is advantageous at the present time that the indicators are
available with reliable values for a large number of galaxies
compared with other indicators,
such as the \HA \ equivalent widths.
The zero-point, i.e., without the present star formation,
is at log~(\fFIRfB) $\simeq$ $-$2, and
\fFIRfB \ becomes larger with increasing of
the present star formation activity
and in this case,
following Devereux and Young (1991) and Sauvage and Thuan (1992),
the present SFR [\Msolar \ yr$^{-1}$] $\simeq$
1.4 $\times 10^{-10}$ \LFIR \ [\Lsolar].
Gallagher et al. (1984) derived a relation between
the $B$ luminosity and the past-averaged SFR,
as SFR [\Msolar \ yr$^{-1}$] $\simeq$
0.3 $\times 10^{-10}$ \LB \ [\Lsolar]
under an assumption that the SFR is constant
during $10^{9}$~yr order and the upper-mass cut-off of 100~\Msolar.
If we use these relations,
the present-to-past averaged SFR (SFR-ratio) is expressed as
$\sim$ (1.4/0.3) ($L_{\rm FIR}/L_{\rm B}$) = (1.4/0.3) (\fFIRfB);
the SFR-ratio = 1,
a constant SFR during $10^{9}$~yr order,
corresponds to log~(\fFIRfB) $\sim$ $-$0.6 to $-$0.7.

\subsection{Distribution of \fFIRfB \ of Galaxies}

\indent

Figure~6 shows a log~(\fFIRfB) versus log~\LB \ relation
for each morphological index
$T$ given in RC3 from
0: S0a to 10: Im.
The blue luminosity is calculated by
\LB \ = 4$\pi D^{2}$\fB,
where $D$ is the distance
obtained by $V_{\rm GSR}/H_{0}$ for
the Hubble constant $H_{0}$ = 75~km~s$^{-1}$~Mpc$^{-1}$ and
$V_{\rm GSR}$ of the recession velocity [km~s$^{-1}$]
with respect to Galactic center (Galaxy Standard of Rest)
taken from RC3.
When $V_{\rm GSR}$ is smaller than 75~km~s$^{-1}$,
we give $D$ = 1~Mpc;
the number of objects with $V_{\rm GSR}$ $<$ 75~km~s$^{-1}$ is 19
and only 1\% of the total sample.
We include one object of $T$ = 11 (cI) into the $T$ = 10 (Im) group.
Circles indicate objects with $IRAS$ detection,
total number is 1306,
including objects without reliable \ften \ but \fFIR \ of which
are estimated by the procedure mentioned in section~3.1.
Crosses show the upper limits of \fFIRfB \
for the objects without $IRAS$ detection;
the total number is 375,
and the true positions are lower than the positions plotted.
The size of the symbols represents the \BT0 magnitude of the object,
as shown at the upper left in each diagram.
Near the left ordinate,
eight small tick marks are shown.
They indicate the detection limits of \fFIRfB,
due to the $IRAS$ detection limits,
for a given \BT0 magnitude,
\BT0 = 16, 15, 14, 13, 12, 11, 10, and 9 mag,
respectively, from upper to lower.
For instance,
an object with \BT0 = 11 is detectable by $IRAS$
if log~(\fFIRfB) $>$ $-2$.

From figure~6,
we confirm not only the earlier claim by others
(e.g., Soifer et al. 1987) that
the values of log~(\fFIRfB) distribute over two orders of magnitude
but also that the wide dispersion of \fFIRfB \
is seen even in each morphological type.
The trend that the distributions of \fFIRfB \ show
no systematic change along the morphological type
is in contrast with the systematic change
of such as \BVT0 color (see figure~1).
If the SFR is constant during $\sim$ $10^{9}$ yr in a galaxy,
log~(\fFIRfB) would be a constant regardless of \LB.
This is not the case in figure~6,
indicating that the star formation activity in the galaxy scale
is variable with a shorter time.
We estimate in appendix~2 that the time scale of the variation
is shorter than about $10^{8}$ yr.

It is an interesting feature that
\fFIRfB \ is independent of \LB \
in early-type spirals (S0a to Sc),
while in late-type spirals (later than Scd, i.e., $T$ $\geq$ 6)
\fFIRfB \ tends to decrease with decreasing \LB;
figure~7 shows log~(\fFIRfB) versus log~\LB \ diagrams
for the 0 $\leq$ $T$ $\leq$ 5 group and
the 6 $\leq$ $T$ $\leq$ 11 group,
respectively.
The positive correlation between \fFIRfB \ and \LB \
seen in the late-type galaxies may be due to the metallicity effect
that less luminous galaxies tend to be metal-poorer
(e.g., Roberts, Haynes 1994).
The 0 $\leq$ $T$ $\leq$ 5 group has few sample with
\LB \ $<$ $10^{9}$\Lsolar \ and the effective span of \LB \
is nearly half of that for the 6 $\leq$ $T$ $\leq$ 11 group.
We need more data of less luminous galaxies
in the 0 $\leq$ $T$ $\leq$ 5 group
in order to examine whether the correlation is
a common feature in both the early- and late-type spiral galaxies.

We make an universal \fFIRfB \ distribution
using the data shown in figure~6.
For more distant galaxies,
only those having higher \fFIRfB \ values are
detectable in the $IRAS$ survey;
we have to correct this ``Malmquist'' bias.
All of our sample galaxies are nearby galaxies
(see table~2) and it is natural to assume no evolution of
the \fFIRfB \ frequency distribution with distance among our sample.
Then,
the \fFIRfB \ frequency distribution of objects
in a small range of $B$ magnitude represents the universal
\fFIRfB \ distribution of galaxies in the Local Universe.
We make \fFIRfB \ distribution histograms
in seven \BT0-magnitude ranges, i.e.,
9 - 10 mag, 10 - 11 mag, 11 - 12 mag, 12 - 13 mag, 13 - 14 mag,
14 - 15 mag, and 15 - 16 mag;
the lower limits of log~(\fFIRfB) used for the analysis are
$-$2.2, $-$1.8, $-$1.4, $-$1.0, $-$0.6, $-$0.2, and 0.2, respectively.
The final frequency histogram is obtained by combining the
magnitude-sliced \fFIRfB \ frequency histograms down to the
detection limit with a weight of number of the objects.
The functions are shown in figure~8 for each morphological type
of 0 $\leq$ $T$ $\leq$ 5.
Figure~8 is the firstly derived universal distribution of
the present star formation activity in spiral galaxies.
The spirals later than $T$ = 6 (Scd) contain the possible metallicity
effect (see figure~7),
and for these galaxies
the indicator of the present star formation activity must be
the values of \fFIRfB \ with some correction of this effect
depending on \LB.
But this is a subject in future work.

The extent of log~(\fFIRfB) is
from about $-$2 to 0.5
for S0a and Sa,
from $-$1.5 to 0.5
for Sab and Sb,
and from $-$1.0 to 0
for Sbc and Sc.
The corresponding present-to-past averaged SFRs range from
almost zero to ten
for S0a and Sa,
from a tenth to ten
for Sab and Sb,
and from several tenths to several
for Sbc and Sc.
The dispersion of log~(\fFIRfB) in each morphological type
is much larger than the difference between the morphological types.
For instance,
the median of log~(\fFIRfB) is $-$0.6 in Sa,
and $-$0.4 in Sb,
and the difference is only 0.2,
while the dispersion is about 2
in both Sa and Sb.
The mean value of the \fFIRfB \ is similar to each other
between morphological types,
which is in contrast with the suggestion by de Jong et al. (1984).
It is interesting to note that
the dispersion of \BVT0 in each morphology is comparable to
the difference of the median value between morphologies
and that the dispersion of \BVT0 in each morphology,
corresponding to $\Delta$ log~(\fBfV) $\sim$ 0.1,
is much smaller than the dispersion of log~(\fFIRfB)
(compare figure~1 with figure~8).
The shape of the distribution histogram of \fFIRfB \ is
different from morphological type to type;
the histogram is Gaussian-like for Sbc and Sc samples,
while the shape is flat for S0a and Sa samples,
indicating the star formation mechanism in S0a and Sa
is different from that in Sbc and Sc.

Kennicutt (1983) derived a similar distribution
of the present-to-past averaged SFRs for about 100 galaxies,
using the \HA \ luminosity,
the total mass estimated using $B$ luminosity,
and the morphology-dependent mass-to-luminosity ratios ($M/L$).
However,
he only pointed out that the preset-to-past averaged SFR
is higher for later-type spirals.
We should emphasize here that
the dispersion of the present-to-past averaged SFR
has an important meaning that
the star formation activity of spiral galaxies,
especially early-type spirals,
has a large time variation.
Coziol and Demers (1995) showed that \LFIR \ has a tight
correlation with \LB \ for the starburst galaxies 
made from Montreal-Blue-Galaxy sample (see their figure~2),
and concluded that the starburst galaxies have
a nearly constant SFR over Gyr scale.
This disagrees with our result.
However, their figure must be seen carefully,
because the luminosity-luminosity relations tend to have a
tight correlation in general.
From our analysis,
the starburst galaxies have high \fFIRfB \ values
i.e., log~(\fFIRfB)~$\sim$ 0
for any \LB \ and morphological types,
and thus necessarily shows a tight correlation
between \LFIR \ and \LB.

\section{Discussion}

\indent

What kind of mechanism causes the large variation
of the present star formation activity
especially for early-type spirals ?
In this section,
we discuss the correlation between the present star formation activity
and the features and environments of galaxies.

\subsection{Bar}

\indent

Whether the bar structure is the effective exciting mechanism
for intense star formation is still an open question.
In our sample data,
85\% of spiral galaxies are divided into
three morphological subgroups concerning the bar structure,
i.e., 
SA (non-barred, 21\% of the total sample),
SX (mixed, 28\%), and
SB (barred, 36\%).
We show in figure~9 the universal frequency histograms of log~(\fFIRfB)
of the barred galaxies (SB sample) for
Sa ($T$ = 1), Sb ($T$ = 3), and Sc ($T$ = 5),
respectively by solid lines.
In each histogram,
the dashed lines indicate the distribution of the non-barred galaxies
(SA sample).
Though the barred galaxy sample has somewhat larger \fFIRfB \
values than the non-barred galaxy sample for the Sa sample,
the distributions of the barred and the non-barred samples are
similar to each other in all three morphologies.
The optical bar structure is, thus, not correlated with
the present star formation activity.
This disagrees with the result by de Jong et al. (1984),
which was derived by using the small number of sample.

\subsection{Nuclear and Disk Starburst}

\indent

It is known that some galaxies show intense starburst
in their nuclear regions.
Devereux (1989) compiled 132 nearby nuclear-region starburst galaxies
according to 10~$\mu$m observations.
Figure~10 shows the log~(\fFIRfB) versus log~\LB \ diagrams for
the nuclear-starburst sample by Devereux (1989)
among our sample for Sa, Sb, and Sc.
The dashed symbols indicate the total sample and
the bold solid circles the nuclear-region starburst sample
among our sample.
It is seen that the nuclear-region starburst galaxies tend to
have the higher values of \fFIRfB \
in all three morphologies.
The existence of the post-nuclear-starburst galaxies
(e.g., Walker et al. 1987; Beckman et al. 1991)
suggests that turning on and off the nuclear-region starburst
often happens.
Kr\"{u}gel and Tutukov (1993) showed 
by an analytic modeling that a possibility of repetitive
starburst in the nuclear region in a time scale of $10^{7-8}$ yr.

Disk star formation also contributes
to the total star formation activity.
Rice et al. (1988) tabulated FIR-to-$B$ diameter-size ratios
for galaxies with large angular sizes;
the ratio measures the development of the star-forming disk.
Figure~11 shows a correlation between the size ratio
and the flux ratio;
filled marks are for the earlier-type galaxies with
$T$ $\leq$ 4 (Sbc) and open marks are 
for the later-type galaxies with $T$ $\geq$ 5 (Sc).
The larger the FIR-to-$B$ flux ratio is,
the larger the FIR-to-$B$ size ratio is;
this trend holds for both earlier and later spirals.
Though the number of the sample is small,
the expansion of the starburst disk seems to be
one of the dominant features of
the high total star formation activity
as well as the nuclear-region starburst.

The \HI \ gas content may be an important factor of
the star formation.
RC3 gives the \HI \ index,
\HI \ index $\equiv$ $-$2.5 log~(\fHIfB) + constant,
where \fHIfB \ is the flux ratio of the \HI \ 21~cm line
and the $B$ band.
Note that the higher value of the \HI \ index indicates
being more \HI-deficient with respect to the $B$ band flux.
RC3 cataloged only reliable \HI \ index values
and for 1204 galaxies among our sample (72~\%).
Figure~12 shows a correlation between log~(\fFIRfB) and
the \HI \ index for the 1204 galaxies.
The circles indicate 962 objects with reliable \fFIRfB,
and the crosses indicate 242 objects without reliable \fFIRfB;
true positions would be more left on the diagram.
For the 962 objects with reliable \fFIRfB,
the gradient is 0.29 or
$\Delta$~log~(\fHIfB)/$\Delta$~log~(\fFIRfB) = 0.29~/~($-$2.5)
$\sim$ $-$0.1
with a correlation coefficient of 0.12;
i.e., there is no correlation between \fHIfB \ and \fFIRfB.
This suggests that the time scale of the present star formation activity
is much shorter than that of gas consumption in a galaxy of
Gyr scale (e.g., Kennicutt et al. 1994).

Kennicutt (1988) and Kennicutt et al. (1989) pointed out that
the more intense star-forming galaxies have more numerous,
more luminous, and larger \HII \ regions.
They suggested that this is due to the variation of
the mass function of the molecular clouds.
M~51 is an Sbc ($T$ = 4.0) galaxy with log~(\fFIRfB) = $-0.14$,
a high value for Sbc sample (see figures~3 and 8 (e)).
Adler et al. (1992) mapped M~51 by \12COJ \ line using
an interferometer
with a spatial resolution of 7$''$ $\times$ 11$''$
(350 $\times$ 550~pc at a distance of 9.6~Mpc),
and found many associations of giant molecular clouds
with sizes of 300 - 800~pc and masses of (1 - 40) $\times$
$10^{7}$ \Msolar,
which are one order of magnitude larger as
compared with the Galactic one.
However,
the high resolution mappings by the molecular lines have been made
for only a few galaxies so far.
High resolution systematic surveys of the molecular gas
in many galaxies are necessary to study
the degree of the gas concentration in the central regions and
the mass functions of the molecular clouds in the disks.

\subsection{Tidal Distortion}

\indent

The galaxy-galaxy interaction is considered to be 
a trigger for the central-region starburst
from observations (e.g., Heisler, Vader 1995)
and from numerical simulations (e.g., Wada, Habe 1992).
We examine our sample galaxies appearing in
catalogs by Arp (1966) and by Vorontsov-Velyaminov (1977)
in which they collected the distorted-shaped galaxies.
Figures~13 and 14 show log~(\fFIRfB) versus log~\LB \ diagrams
for the Arp sample and the Vorontsov-Velyaminov sample,
respectively,
in the same way as figure~10.
For Sb galaxies,
both the distorted-shaped samples tend to show high \fFIRfB \ values.
However, for Sa and Sc galaxies,
the distorted-shaped samples show no different distribution
from that of the total sample.
60\% of the Arp sample shown in figure~13 and 68\% of
the Vorontsov-Velyaminov sample shown in figure~14
share the common sample.
The common sample has also a similar behavior on log~(\fFIRfB) versus
log~\LB \ diagram.
Galaxies with distorted shape, thus,
does not always indicate the higher
present star formation activity.

\subsection{Spatial Number Density of Galaxies}

\indent

The spatial number density of galaxies is considered to
affect the star formation history of the galaxies.
We investigate the galaxies in the three extreme environments,
which are field isolated galaxies,
compact group members, and
cluster members.
We refer to the isolated sample catalog by Karachentseva (1973).
Figure~15 shows a log~(\fFIRfB) versus log~\LB \ diagram
in the same way as figure~10.
The number of the Karachentseva sample for Sa is only two.
For Sb and Sc galaxies,
the distribution of the Karachentseva sample is similar to
that of the total sample.
It seems, therefore,
that the environment of ``extreme field'' does not
affect the star formation activity.

Hickson (1982, 1993) made the Compact Group catalog.
The compact groups give the most highest spatial number density
environment in 10~kpc scale, 
though the group itself is isolated in Mpc scale.
We have omitted from our sample the companion systems that
we can not identify the association of $IRAS$ point sources
on such individual objects,
as mentioned in section~2.1,
but some galaxies in relatively loose compact groups are left
in our sample.
Figure~16 shows a log~(\fFIRfB) versus log~\LB \ diagram
for the Hickson sample in the same way as figure~10.
Though the number of the sample is small,
it seems that there is no significant difference on this diagram
between the group member sample and the total sample.

For the cluster member sample,
we refer to the rich cluster catalog
by Abell et al. (1989).
We pick up cluster members by satisfying
both of two criteria that
(1) the location is within 0.5 Abell radius
(1~$R_{\rm A} \equiv 1.7 z^{-1}$ arcmin,
$z$ is the redshift of the cluster)
from the center of the cluster, and that
(2) the radial velocity difference is within 500~km~s$^{-1}$ with
respect to the mean cluster radial velocity.
This procedure picks up effectively galaxies
in the central region of the rich clusters. 
Since our sample galaxies are mostly nearby objects (see table~2),
they contain a small number of Abell cluster members.
We add three nearby poor clusters,
Virgo, Cancer, and Pegasus~I clusters as the cluster sample;
the criteria of the membership are the same as for 
the Abell cluster members.
Figure~17 shows a log~(\fFIRfB) versus log~\LB \ diagram
in the same way as figure~10.
Though the number of the cluster member is small,
again it seems that there is no significant difference on the diagram
between the cluster member sample and the total sample.
Thus,
the spatial number density of galaxies does not correlate with
the activity of the present star formation
of spiral galaxies.

\section{Summary}

\indent

We constructed a new large data set of 1681 nearby spiral galaxies
from available databases,
and studied the variation of the present star formation activity.
Using these data and the FIR and $B$-band surface intensity image data
of M~31,
we showed that \fFIRfB \ is a useful indicator of
the present star formation activity in spiral galaxies;
log~(\fFIRfB) $\simeq$ $-$2 corresponds to
no present star formation activity,
where all FIR emission is due to the cirrus component,
and \fFIRfB \ increases with the present star formation activity.
The value of log~(\fFIRfB) would be about $-$0.7 to $-$0.6 in
the case that the present SFR is the same as the past-averaged SFR
and the maximum value of log~(\fFIRfB) amounts to about 0.5.
Analyzing the distribution of \fFIRfB \ of the sample galaxies,
we showed that the present star formation activity
has a great variety even within the same morphological type
of galaxies
and that the star formation activity
changes discontinuously in a short time scale.
We also showed that the intense star formation activity
does not continue more than $10^{8}$~yr.
For spirals with $T$ = 4 to 5,
the time variation of the present star formation activity
is not so large as that for spirals with $T$ = 0 to 1.
We have made a universal frequency histogram of log~(\fFIRfB),
the present star formation activity indicator,
for each morphological type of spiral galaxies.
We pointed out that the shape of the universal histogram of
log~(\fFIRfB) in Sbc and Sc sample is Gaussian-like,
while that in S0a and Sa sample is fairly flat.
The systematic difference of the log~(\fFIRfB) distribution
along the morphological type suggests
that the mechanisms regulating the present star formation
activity are internal processes in galaxies;
the different morphological type may cause the different star formation
mechanism and
the different star formation mechanism may maintain the morphology.

We investigated what structures and environments of galaxies 
are correlated with the present
star formation activity.
The nuclear-region starburst may be a state of
the intense star formation activity in the whole galaxy.
The disk-wide starburst also has an important contribution to 
the total star formation activity.
The environmental effects are little correlated with 
the present star formation activity,
though the galaxy-galaxy interaction is considered to be a
triggering mechanism for the intense star formation in galaxies.
This is consistent with the suggestion mentioned above that the
intense star formation does not continue for a long time.

\bigskip
\bigskip

We would like to thank 
Keiichi Kodaira for critical reading of the manuscript and
Hidemi Suzuki for kindly allowing us to use the $B$ band image of M~31.
We made use of NDADS/ARMS on-line service
and thank Walter Rice, Melissa Larkin, and Seth Digel
for their kind guide.
This research has also made use of
the NASA/IPAC Extragalactic Database (NED)
which is operated by the Jet Propulsion Laboratory, Caltech,
under contact with the National Aeronautics and Space Administration.

\newpage

\section*{Appendix 1.
Relation of log~(\fFIRfB) vs Inclination of Galaxies}

\indent
The \BT0-system in RC3 corrects statistically the internal
absorption in galaxy depending on apparent inclination of galaxy.
We investigate the distribution of log~(\fFIRfB) as a function
of apparent inclination of galaxy to check whether the
distribution of log~(\fFIRfB) is seriously affected by the
extinction correction.
Apparent inclinations of galaxies versus log~(\fFIRfB) diagrams
are shown for Sa ($T$ = 1), Sb ($T$ = 3), and Sc ($T$ = 5)
samples in figure~18 (a) and for the total 1681 sample in
figure~18 (b).
We obtain the apparent inclination of a galaxy, $i$ [degree],
from apparent ratio of minor-to-major axes at a surface brightness
of 25~mag~arcsec$^{-1}$ at $B$ band, $R$,
using an equation shown in {\it Nearby Galaxies Catalog}
by Tully (1988),
${\rm cos}^{2}(i - 3^{\rm o}) = (R^{2} - 0.2^{2})/(1 - 0.2^{2})$.
It is found from figure~18 that the mean value and the dispersion
of log~(\fFIRfB) are independent of the inclination.

\section*{Appendix 2.
The Time Scale of the Intense Star Formation Activity}

\indent
The actively star-forming galaxies have values of log~(\fFIRfB)
as high as 0 to 0.5.
If galaxies would maintain such high activities for a long time,
B-type and A-type stars are accumulated in a large number;
the plots on the log~(\fFIRfB) versus log~\LB \ diagrams
shown in figures~6 and 7 would move
to the lower right direction along the line with a coefficient
of $-$1.

We estimate the evolution of \fFIRfB \ and \LB \
under an assumption of the constant star formation rate
at a high level.
We assume the IMF of Salpeter (1955),
i.e., $n(m) dm \propto m^{-2.35} dm$,
where $n(m) dm$ is the number of stars having masses
of $m$ to $m + dm$.
The mass versus bolometric luminosity relation of stars is
$l(m) \propto m^{3.45}$ (Allen 1973).
The total luminosity of stars having masses of $m$ to $m + dm$ is,
therefore, $n(m)l(m) dm \propto m^{1.10} dm$.
The life time of a star is written as
$t(m) \propto m/l(m) \propto m^{-2.45}$,
or $t(m) = 10^{10} m^{-2.45}$~yr,
for $m$ in the unit of \Msolar.
The starburst forms stars with masses from the lowest one of
$m_{l}$ to the most massive one of $m_{u}$.
Let $t_{0}$ be the time since the beginning
of the constant starburst,
the resultant total luminosity relative to the luminosity by a
single-generation starburst,
the time of which is put here by a life time of the most massive star,
is
$L/L_{\rm sing} \equiv (\int_{m_{0}}^{m_{u}} m^{1.10} m^{-2.45} dm +
m_{0}^{-2.45} \int_{m_{l}}^{m_{0}} m^{1.10} dm)/
(m_{u}^{-2.45} \int_{m_{l}}^{m_{u}} m^{1.10} dm)$,
where $m_{0}$ is a mass of the most massive star
which has survived for the starburst duration, $t_{0}$,
and usually $m_{u} > m_{0} \gg m_{l}$.
Note that stars with masses around $m_{0}$ contribute
the most to the total luminosity.
We also assume that the FIR luminosity generated
by the recent starburst
dominates the observed FIR luminosity because of
very intense present star formation activity.
By such a continuous intense star formation,
the total luminosity $L$ increases with time,
and $L/L_{\rm sing}$ would be about ten when $t_{0}$ $\sim$ $10^{8}$~yr
(corresponding to $m_{0}$ $\sim$ 6~\Msolar)
for $m_{u}$ = 10 to 50~\Msolar.
In this time,
$L$ nearly corresponds to \LB \ because of blue colors with
$m_{0}$ $\sim$ 6~\Msolar \ and $L_{\rm sing}$ to \LFIR;
i.e.,
\LB \ has increased by a factor of ten and
\fFIRfB \ has decreased by a factor of ten.
There would be, therefore,
a ``forbidden'' region in the upper right corner of the 
log~(\fFIRfB) versus log~\LB \ distribution of galaxies.
This is not the case in figures~6 and 7 and thus
the intense star formation activity of a galaxy continues only
shorter than $10^{8}$~yr.

\newpage

\section*{References}

\parindent=-0.7cm
\leftskip=0.7cm

Abell G.O., Corwin H.G., Jr., Olowin R.P. 1989,
ApJS, 70, 1

{\it
ADC CD-ROM Selected Astronomical Catalogs Volume 1
}
prepared by L.E. Brotaman, S.E. Gessner,
NASA/GSFC Astronomical Data Center

Adler D.S., Lo K.Y., Wright M.C.H., Rydbeck G., Plante R.L.,
Allen R.J. 1992,
ApJ, 392, 497

Allen C.W. 1973,
{\it
Astrophysical Quantities Third Edition
}
(The Athlone Press, University of London, London)

Arimoto N., Yoshii Y. 1986,
A\&A, 164, 260

Arp H.C. 1966,
ApJS, 14, 1

Beckman J.E., Varela A.M., Mu\~{n}oz-Tu\~{n}on C.,
V\'{i}lchez J.M., Cepa J. 1991,
A\&A, 245, 436

Beichman C.A. 1987,
ARA\&A, 25, 521

Beichman C.A., Myers P.C., Emerson J.P., Harris S., Mathieu R.,
Benson P.J., Jennings R.E. 1986,
ApJ, 307, 337

Belfort P., Mochkovitch R., Dennefeld M. 1987,
A\&A, 176, 1

Bothun G.D., Lonsdale C.J., Rice W. 1989,
ApJ, 341, 129

Bruzual A.G. 1983,
ApJ, 273, 105

Campos-Aguilar A., Moles M. 1991,
A\&A, 241, 358

{\it
Cataloged Galaxies and Quasars Observed in the IRAS Survey
}
1985,
prepared by C.J. Lonsdale, G. Helou, J. Good, W. Rice
(JPL, Pasadena)
(CGQIRAS)

Coziol R., Demers S. 1995,
ApL\&Comm, 31, 41

de Jong T., Clegg P.E., Soifer B.T., Rowan-Robinson M., Habing H.J.,
Houck J.R., Aumann H.H., Raimond E. 1984,
ApJL, 278, L67

Devereux N.A. 1989,
ApJ, 346, 126

Devereux N.A., Jacoby G., Ciardullo R. 1995,
AJ, 110, 1115

Devereux N.A., Price R., Wells L.A., Duric N. 1994,
AJ, 108, 1667

Devereux N.A., Scowen P.A. 1994,
AJ, 108, 1244

Devereux N.A., Young J.S. 1990,
ApJL, 350, L25

Devereux N.A., Young J.S. 1991,
ApJ, 371, 515

Dressler A., Thompson I.B., Shectman S.A. 1985,
ApJ, 288, 481

Fabbiano G., Gioia I.M., Trinchieri G. 1988,
ApJ, 324, 749

Gallagher J.S.,III, Hunter D.A., Tutukov A.V. 1984,
ApJ, 284, 544

Gavazzi G., Boselli A., Kennicutt R. 1991,
AJ, 101, 1207

Gavazzi G., Jaffe W. 1985,
ApJL, 294, L89

Guiderdoni G., Rocca-Volmerange B. 1987,
A\&A, 186, 1

Hamabe M., Ichikawa S. 1992,
in Proc of Astronomical Data Analysis Software and System I,
ASPC, 25, ed D. Worrall et al.
(ASP, San Francisco) p325

Heisler C.A., Vader J.P. 1995,
AJ, 110, 87

Hickson P. 1982,
ApJ, 255, 382

Hickson P. 1993,
ApL\&Comm, 29, 1

{\it
IRAS Faint Source Catalog
$|b| > 10$ Degrees Version 2.0
}
1989,
prepared by M. Moshir 
(GPO, Washington, DC)
(FSC)

{\it
IRAS Point Source Catalog
}
1985,
Joint $IRAS$ Science Working Group
(GPO, Washington, DC)
(PSC)

{\it
IRAS Small Scale Structures Catalog
}
1986,
prepared by G. Helou, D.W. Walker
(GPO, Washington, DC)
(SSSC)

Karachentseva V.E. 1973,
Soob. Spets. Astrofiz. Obs. 
(Communication of the Spetial Astrophysical Observatory), No.8

Keel W.C. 1993,
AJ, 106, 1771

Kenney J.D.P., Wilson C.D., Scoville N.Z., Devereux N.A.,
Young J.S. 1992,
ApJL, 395, L79

Kennicutt R.C.,Jr. 1988,
ApJ, 334, 144

Kennicutt R.C.,Jr. 1983,
ApJ, 272, 54

Kennicutt R.C.,Jr. 1992,
ApJ, 388, 310

Kennicutt R.C.,Jr., Edgar B.K., Hodge P.W. 1989,
ApJ, 337, 761

Kennicutt R.C.,Jr., Kent S.M. 1983,
AJ, 88, 1094

Kennicutt R.C.,Jr., Tamblyn P., Congdon C.W. 1994,
ApJ, 435, 22

Kr\"{u}gel E., Tutukov A.V. 1993,
A\&A, 275, 416

Larson R.B., Tinsley B.M. 1978,
ApJ, 219, 46

Moss C., Whittle M. 1993,
ApJL, 407, L17

{\it
NAO/ADAC CD-ROM Catalog Volume~1.
Galaxies and Non-stellar Objects
}
1994,
prepared by M. Hamabe, S. Ichikawa, S. Nishimura

{\it
Nearby Galaxies Catalog
}
1988,
Tully R.B.
(Cambridge Universit Press, Cambridge)

Persson C.J.L., Helou G. 1987,
ApJ, 314, 513

Rice W. 1993,
AJ, 105, 67
({\it IRAS HiRes})

Rice W., Lonsdale C.J., Soifer B.T., Neugebauer G.,
Kopan E.L., Lloyd L.A., de Jong T., Habing H.J. 1988,
ApJS, 68, 91

Roberts M.S., Haynes M.P. 1994,
ARA\&A, 32, 115

Roberts W.W.,Jr., Huntley J.M., van Albada G.D. 1979,
ApJ, 233, 67

Salpeter E.E. 1955,
ApJ, 121, 161

Sandage A. 1986,
A\&A, 161, 89

Sauvage M., Thuan T.X. 1992,
ApJL, 396, L69

Sauvage M., Thuan T.X. 1994,
ApJ, 429, 153

Searle L., Sargent W.L.W., Baguolo W.G. 1973,
ApJ, 179, 427

Shu F.H., Adams F.C., Lizano S. 1987,
ARA\&A, 25, 23

Soifer B.T., Boehmer L., Neugebauer G., Sanders D.B. 1989,
AJ, 98, 766

Soifer B.T., Sanders D.B., Madore B.F., Neugebauer G.,
Danielson G.E., Elias J.H., Lonsdale C.J., Rice W.L. 1987,
ApJ, 320, 238

{\it
Third Reference Catalogue of Bright Galaxies
}
1991,
de Vaucouleurs G., de Vaucouleurs A., Corwin H.G.,Jr., 
Buta R.J., Paturel G., Forqu\'{e} P.
(Springer-Verlag, New York)

Vorontsov-Velyaminov B.A. 1977,
A\&AS, 28, 1

Wada K., Habe A. 1992,
MNRAS, 258, 82

Walker C.E., Lebofsky M.J., Rieke G.H. 1988,
ApJ, 325, 687

Wood D.O.S., Churchwell E. 1989,
ApJ, 340, 265

Xu C. 1990,
ApJL, 365, L47

\newpage

\parskip=0.2cm

Figure Captions
\bigskip

Figure~1.~
Frequency histogram of \BVT0 of galaxies
for each morphological type.
The morphological type and index are shown at the right of
each diagram.
The later the morphological type is,
the bluer the color is.
The dispersion of the color in each morphological type
is small,
corresponding to $\Delta$ log~(\fBfV) $\sim$ 0.1.

Figure~2.~
A log~(\ftenfsix) versus log~(\fsixfB) diagram
for 1150 sample galaxies with reliable \fsix \ and \ften \ values.
A straight line is a regression line and the correlation
coefficient is 0.72.
Most of the objects showing the large deviation from the regression
line are irregular galaxies.
Using only objects with 0 $\geq$ $T$ $\geq$ 5,
the correlation coefficient is 0.80.
Using the relation of the regression line,
log~(\ftenfsix) = $-$0.23 log~(\fsix \ [Jy] / \fB \ [W~m$^{-2}$])
+ 3.41,
we estimate \ften \ for 156 objects with upper-limited
\ften \ and reliable \fsix.

Figure~3.~
A log~(\ftenfsix) versus log~(\fFIRfB) diagram
for 1150 sample galaxies.
Large circles indicate 916 early-type spirals
(0 $\geq$ $T$ $\geq$ 5)
and small circles indicate 473 late-type spirals
(6 $\geq$ $T$ $\geq$ 11).
Six famous spiral galaxies without strong AGN
are denoted by large filled
squares with numbers;
1: M~31 ($T$ = 3),
2: M~81 ($T$ = 2),
3: M~33 ($T$ = 6),
4: M~101 ($T$ = 6),
5: M~51 ($T$ = 4),
6: M~83 ($T$ = 5).
The three areas enclosed by solid lines show the color-color relation
corresponding to
(A) the \HII \ regions,
(B) the non-\HII \ regions, and
(C) the nuclear regions
in M~31,
as shown in figure~5.
The composition of the \HII \ regions (A) and the non-\HII \ regions (B)
in various fraction makes the sequence shown by the bold solid line
(see text),
which is similar to the sequence of the galaxy distribution
on this diagram.

Figure~4.~
(a) The $B$-band image taken at Ouda Station, Kyoto University.
Foreground stars are removed using IRAF
task of `imedit'.
The north is top and the east is left.
The image size is 1.$'$68 $\times$ 2.$'$00.
(b) The \fFIRfB \ image made by the $B$-band image shown in (a)
and $IRAS$ $HiRes$ image data of Rice (1993).
The sample locations for measuring the surface intensities
in FIR and $B$ bands are indicated by circles.
The size of the circles
corresponds to the aperture size for the photometry,
0.$'$5 in radius.
The numbers 1 - 15 represent \HII \ regions,
16 - 25 non-\HII \ regions in disk,
26 - 32 bulge regions, and
33 - 35 central regions.

Figure~5.~
The same as figure~3 but for various areas in M~31
with an aperture radius of 0.$'$5, i.e.,
15 points from \HII \ regions in disk (2 to 8~kpc from the nucleus),
10 points from non-\HII \ regions in disk (2 to 8~kpc),
7 points from bulge region (1 to 2~kpc), and
3 points from nuclear region (within 0.3~kpc),
the locations of which are shown in figure~4.
The meanings of the symbols are shown in the diagram.
The solid line (A) encloses the \HII \ regions,
the solid line (B) encloses both the non-\HII \ regions in disk
and bulge regions,
which occupy the same area and are both called as
non-\HII \ regions.
The solid line (C) encloses the three points in the central 0.3~kpc.
The big star mark is the position of the whole M~31,
which is inside the region~B,
indicating that most of the FIR emission of M~31
comes from cirrus component.

Figure~6.~
A log~(\fFIRfB) versus log~\LB \ diagram for each morphological type,
from (a) $T$ = 0 (S0a)
to (k) $T$ = 10, 11 (Im/cI).
Circles indicate objects with measured \fFIR,
while crosses indicate ones with upper-limited \fFIR.
Size of the symbols shows \BT0 magnitude
as shown in the upper left corner of the diagram.
Note that \LB \ is written in the unit of solar luminosity;
\Lsolar = 3.85 $\times$ $10^{26}$ W.
Small tick marks near the left ordinate show the detection limit
of \fFIRfB \,
due to the $IRAS$ detection limit,
for given apparent \BT0 magnitude,
\BT0 = 16, 15, 14, 13, 12, 11, 10, and 9 mag,
respectively from upper to lower.

Figure~7.~
The same as figure~6 but for
the 0 $\leq$ $T$ $\leq$ 5 group and the 6 $\leq$ $T$ $\leq$ 11 group.
For the early-type spirals,
\fFIRfB \ distributes without \LB-dependence,
while for the late-type spirals,
\fFIRfB \ tends to increase with increasing \LB.
Most of the upper limit sample in the 6 $\leq$ $T$ $\leq$ 11 group
are irregulars;
see figure~6 (k).

Figure~8.~
A universal frequency histogram of log~(\fFIRfB)
for each morphological type.
From upper to lower panels,
(a) $T$ = 0 (S0a), (b) $T$ = 1 (Sa), (c) $T$ = 2 (Sab),
(d) $T$ = 3 (Sb), (e) $T$ = 4 (Sbc), (f) $T$ = 5 (Sc).
We show only objects earlier than Sc,
because for the later-type galaxies,
\fFIRfB \ depends on \LB.
The dashed line indicates the upper limit value.
log~(\fFIRfB) $\simeq$ $-$2 corresponds to
no present star formation activity.
The histograms are Gaussian-like for Sbc and Sc samples,
while they are rather flat for S0a and Sa samples.

Figure~9.~
A universal frequency histogram of log~(\fFIRfB)
for the barred (SB) sample
(solid line) and
the non-barred (SA) sample (dashed line).
Only Sa ($T$ = 1), Sb ($T$ = 3), and Sc ($T$ = 5) samples
are shown here.
The number of the barred sample is
65, 117, and 78 for $T$ = 1, 3, and 5, respectively, and
the number of the non-barred sample is
38, 55, and 84, respectively.
There is no significant difference between both samples.

Figure~10.~
A log~(\fFIRfB) versus log~\LB \ diagrams
for the nuclear-region starburst
galaxies listed in Devereux (1989).
Size of the symbols shows \BT0 magnitude as shown in the upper
left corner of the diagram and the magnitude ranges are the same
as those in figure~6.
The bold solid symbols indicate the nuclear-region starburst sample.
The dashed symbols indicate the total sample,
the same as shown in figure~6.
Generally the nuclear-starburst sample has high \fFIRfB \ values.

Figure~11.~
A log~(\fFIRfB) versus  FIR-to-B size ratio diagram for 
the galaxies listed in Rice et al. (1988).
Filled circles are for the earlier-type galaxies with
$T$ $\leq$ 4 (Sbc) and open circles are for the later-type galaxies
with $T$ $\geq$ 5 (Sc).
The more the FIR-disk develops,
the larger the star formation indicator,
\fFIRfB, is.

Figure~12.~
A log~(\fFIRfB) versus the \HI \ index diagram for 1204 galaxies
with the \HI \ index values cataloged in RC3.
The circles indicate the objects with reliable \fFIRfB,
and the crosses indicate the objects without reliable \fFIRfB;
true positions would be more left on the diagram.
There is no correlation between log~(\fFIRfB) and the \HI \ index.

Figure~13.~
The same as figure~10 but for the peculiar-shaped
galaxies listed in Arp (1966).
In Sb panel,
the Arp sample shows a high \fFIRfB \ value
compared with the total sample,
however,
in Sa and Sc panels,
the Arp sample does not show any peculiar distribution on the diagrams.

Figure~14.~
The same as figure~10 but for the peculiar-shaped
galaxies listed in Vorontsov-Velyaminov (1977).
The trend is almost the same as that for the Arp sample
in figure~13.

Figure~15.~
The same as figure~10 but for the field isolated
galaxies listed in Karachentseva (1973).
The number of the Karachentseva sample is only two for Sa.
For Sb and Sc,
there is no significant difference of distribution
on the diagram between the Karachentseva and the total samples.

Figure~16.~
The same as figure~10 but for the compact group members
listed in Hickson (1982, 1993).
Though the number of group members is small,
it seems that they do not have different distribution
from that of the total sample on the diagram.

Figure~17.~
The same as figure~10 but for the cluster members.
We take into account rich clusters cataloged by Abell et al. (1989)
and three poor clusters, Virgo, Pegasus~I, and Cancer. 
Though the number of cluster members are small,
it seems that they do not have different distribution
from that of the total sample.

Figure~18.~
Distributions of log~(\fFIRfB) as a function of
apparent inclinations of galaxies
(a) for Sa ($T$ = 1), Sb ($T$= 3),
and Sc ($T$ = 5) samples and
(b) for the total 1681 sample.
Open circles are for galaxies with reliable \fFIRfB \ and
crosses indicate the upper limits of \fFIRfB.
Inclinations of 0\deg \ and 90\deg \ mean face-on and edge-on,
respectively.

\newpage

\parskip=0cm

Table 1. The morphology and magnitude distribution of the sample.

\begin{table*}[h]
 \begin{tabular}{rcrrrrrrrrcr}
 \hline \hline
 $T^{*}$ &     & \multicolumn{8}{c}{\BT0$^{\dag}$} && Total \\
         & \ \ & - 9 & 9 - 10 & 10 - 11 & 11 - 12 & 12 - 13 &
 13 - 14 & 14 - 15 & 15 - & \ \ & \\
 \hline
  0 \ &&  0 &  1 &  6 & 11 & 40 & 25 & 13 &  5 && 101 \\
  1 \ &&  1 &  1 &  4 & 25 & 45 & 52 & 15 &  9 && 152 \\
  2 \ &&  3 &  3 & 12 & 20 & 43 & 24 & 15 &  1 && 121 \\
  3 \ &&  1 & 13 & 21 & 43 & 77 & 70 & 24 & 13 && 262 \\
  4 \ &&  2 &  4 & 25 & 60 & 72 & 67 & 20 &  7 && 257 \\
  5 \ &&  1 &  4 & 27 & 58 &103 & 94 & 24 &  4 && 315 \\
  6 \ &&  4 &  5 & 11 & 32 & 29 & 27 & 18 &  3 && 129 \\
  7 \ &&  3 &  5 &  3 & 15 & 13 & 10 & 11 &  1 &&  61 \\
  8 \ &&  0 &  1 &  0 &  7 & 15 & 12 & 14 &  1 &&  50 \\
  9 \ &&  3 &  0 &  3 & 10 & 22 & 22 & 27 &  3 &&  90 \\
 10 \ &&  0 &  2 &  5 &  8 & 20 & 41 & 55 & 11 && 142 \\
 11$^{\S}$
      &&  0 &  0 &  0 &  0 &  0 &  0 &  1 &  0 &&   1 \\
 \hline
 Total&& 18 & 39 &117 &289 &479 &444 &237 & 58 &&1681 \\
 \hline
 \end{tabular}
\end{table*}

$^{*}$~
The morphological index given in RC3; i.e.,
0:S0a, 1:Sa, 2:Sab, 3:Sb, 4:Sbc, 5:Sc, 6:Scd, 7:Sd, 8:Sdm,
9:Sm, 10:Im, 11:cI.

$^{\dag}$~
The classes of \BT0 magnitude range;
- 9 means \BT0 $<$ 9.0, 9 - 10 means 9.0 $\leq$ \BT0 $<$ 10.0,
and 15 - means \BT0 $\geq$ 15.0.

$^{\S}$~
One object of $T$ = 11 is combined into the $T$ = 10 group
in all analyses in this paper.

\newpage

Table 2. The morphology and radial velocity distribution of the sample.

\begin{table*}[h]
 \begin{tabular}{rcrrrrrrrrrcr}
 \hline \hline
 $T^{*}$ && \multicolumn{9}{c}{$V_{\rm GSR}$ [km s$^{-1}$]$^{\dag}$} && Total \\
         && $<$ 1000 & 1000 & 2000 & 3000 & 4000 &
 5000 & 6000 & 7000 & 8000 - & \ \ & \\
 \hline
  0 \ && 10 & 32 & 18 &  7 & 10 & 11 &  6 &  3 &  4 && 101 \\
  1 \ && 16 & 39 & 22 & 12 & 20 & 21 &  6 &  3 & 13 && 152 \\
  2 \ && 15 & 29 & 12 & 19 & 15 & 14 &  4 &  6 &  7 && 121 \\
  3 \ && 20 & 51 & 45 & 32 & 38 & 25 & 17 &  3 & 31 && 262 \\
  4 \ && 19 & 57 & 43 & 30 & 25 & 25 & 18 & 11 & 29 && 257 \\
  5 \ && 29 & 66 & 59 & 24 & 47 & 42 & 13 & 12 & 23 && 315 \\
  6 \ && 32 & 29 & 19 & 12 & 12 & 11 &  3 &  3 &  8 && 129 \\
  7 \ && 25 & 20 &  5 &  3 &  6 &  0 &  1 &  1 &  0 &&  61 \\
  8 \ && 11 & 20 &  6 &  6 &  5 &  0 &  1 &  1 &  0 &&  50 \\
  9 \ && 37 & 36 & 13 &  2 &  1 &  1 &  0 &  0 &  0 &&  90 \\
 10 \ && 78 & 34 & 15 &  9 &  3 &  0 &  0 &  0 &  2 && 142 \\
 11$^{\S}$
      &&  0 &  0 &  0 &  0 &  0 &  0 &  0 &  0 &  1 &&   1 \\
 \hline
 Total&&292 &413 &257 &156 &182 &150 & 70 & 43 &118 &&1681 \\
 \hline
 \end{tabular}
\end{table*}

$^{*}$~
The morphological index given in RC3; i.e.,
0:S0a, 1:Sa, 2:Sab, 3:Sb, 4:Sbc, 5:Sc, 6:Scd, 7:Sd, 8:Sdm,
9:Sm, 10:Im, 11:cI.

$^{\dag}$~
The radial velocity $V_{\rm GSR}$ [km~s$^{-1}$]
system given in RC3 is adopted.
$<$ 1000 means $V_{\rm GSR}$ $<$ 1000,
1000 means 1000 $\leq$ $V_{\rm GSR}$ $<$ 2000,
and 8000 - means $V_{\rm GSR}$ $\geq$ 8000.

$^{\S}$~
One object of $T$ = 11 is combined into the $T$ = 10 group
in all analyses in this paper.

\end{document}